\documentclass[aps,pra,twocolumn,groupedaddress,10pt,floatfix,showpacs]{revtex4-1}
\usepackage{amsmath}
\usepackage{amsfonts}
\usepackage{amssymb}

\usepackage{hyperref}
\hypersetup{
    colorlinks=true,
    linkcolor=black,
    citecolor=black,
    filecolor=black,
    urlcolor=black,
}

\usepackage[normalem]{ulem} %For strikeout: \sout{}
\usepackage{color}		% For textcolor{red}{}
\usepackage{graphicx}   % For inculdefigure
\graphicspath{{figures/arxiv/}}		% You need to specify either figures/aps/ or figures/arxiv/

\newcommand{\THz}{\,\text{THz}} 
\newcommand{\micron}{\,\mu\text{m}} 
\newcommand{\WOverCmSqr}{\,\text{W}/\text{cm}^2} 
\newcommand{\nitrogen}{$\text{N}_2$}

\newcommand{\abs}[1]{\left| #1 \right|} 						% for absolute value
\newcommand{\pd}[2]{\frac{\partial #1}{\partial #2}} 			% for partial derivatives
\newcommand{\pdd}[2]{\frac{\partial^2 #1}{\partial #2^2}} 		% for double partial derivatives
\let\baraccent=\= 												% rename builtin command \= to \baraccent
\renewcommand{\=}[1]{\stackrel{#1}{=}} 							% for putting numbers above =
\newcommand{\unit}[1]{\ensuremath{\, \mathrm{#1}}}				% Make sure units have proper formatting
\newcommand{\sinc}[0]{\text{sinc} }                  	% sinc(x) = sin(x)/x
\begin{document}

\title{THz generation by optical Cherenkov emission from ionizing two-color laser pulses}

\author{L. A. Johnson}
\author{J. P. Palastro} 
\author{T. M. Antonsen} 
\author{K. Y. Kim} 
\affiliation{University of Maryland, College Park, Maryland 20742, USA}

\date{\today}

\begin{abstract}
Two-color photoionization produces a cycle-averaged current driving broadband, conically emitted THz radiation. 
We investigate, through simulation, the processes determining the angle of conical emission. 
We find that the emission angle is determined by an optical Cherenkov effect, where the front velocity of the current source is faster than the THz phase velocity.
\end{abstract}

% Nonlinear Optics, Filamentation in plasma,   Ionization of atoms field, Frequency conversion (nonlinear optics), 
%\keywords{}

\maketitle

\section{Introduction}

Ultrashort, ultraintense laser pulses propagating through and ionizing gases have produced intense pulses of THz radiation. 
The large electric and magnetic fields of these pulses are potentially useful for a variety of applications \cite{Sherwin2004}. 
For example, intense magnetic fields ($\approx 1~\text{T}$) with subpicosecond duration can be used for coherent control of the spin degree of freedom, in spintronic systems, exciting and deexciting spin waves \cite{Kampfrath2010}. 
In molecular spectroscopy, the high electric fields ($\approx 1~\text{MV/cm}$) of THz pulses can be used to orient molecules for transient birefringence and free induction decay measurements \cite{Fleischer2011}.  
Using ultrashort laser pulses to generate THz via air breakdown may provide a scalable, compact source of few-cycle THz pulses when compared to modern accelerators \cite{Sherwin2004}. 
Scaling to higher energies is possible because field-induced breakdown of the medium is a feature, not a limitation. 
In addition, the compact nature of these sources and their ability to use air as a generation medium potentially allows for standoff capabilities \cite{Daigle2012}.
Generating the THz close to its target decreases the distance over which the THz must propagate, limiting atmospheric absorption \cite{Lu2006}.
Developing such a THz source will require an understanding of the competing nonlinear interactions in atmospheric gases.

Cook \textit{et al}.\ \cite{Cook2000} reported using an ultrashort laser pulse consisting of two colors, a fundamental ($800~\text{nm}$) and its second harmonic ($400~\text{nm}$), to produce approximately $5~\text{pJ}$ of THz radiation between $0$ and $5\THz$. Recent experiments have been able to reach $7~\mu\text{J}$ for frequencies below $10\THz$ \cite{Oh2013}. 
The generation mechanism was originally explained as optical rectification via an unspecified third-order nonlinearity. 
In 2007, Kim \textit{et al}.\  \cite{Kim2007,Kim2009} described the process as tunneling ionization that induces transverse currents on the time scale of the laser pulse envelope ($50~\text{fs}$). 
Recent three-dimensional simulations by Berg\'{e} \textit{et al}.\ \cite{Berge2013} have shown that the bulk of the THz generation in argon, which has a similar ionization potential to {\nitrogen}, can be explained by this mechanism. 
One feature in recent experiments \cite{You2012} is that the THz radiation is observed to emerge in the forward direction (parallel to the axis of the two laser pulses) in a cone with angle roughly $4^\circ - 7^\circ$ with respect to the optical axis.
In this paper we will explore the mechanism contributing to this effect.

\begin{figure*}
	\centering	
    \includegraphics[trim={0.5in 0 -0.5in 0},clip,width=6.0in]{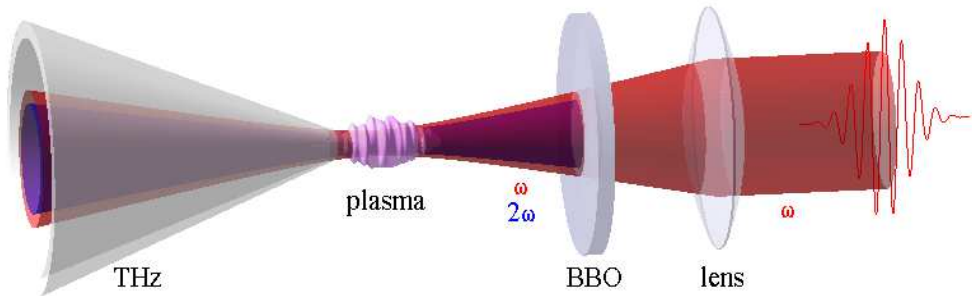}
    \caption{This is a schematic of the experimental setup that is being simulated. The simulation domain includes everything to the left of the BBO crystal. As the two-color pulse (red and blue) approaches focus, it ionizes the gas and generates a plasma. The THz radiation (gray)  exits the other side of the plasma as a cone \cite{You2012}. 
    \label{fig: exp setup}}    
\end{figure*}

We are interested in modeling an experimental setup similar to that of You \textit{et al}.\ \cite{You2012}, as shown in Fig.\ \ref{fig: exp setup}. 
In our setup, an ultrashort pulse with a wavelength of $800~\text{nm}$, duration of $25~\text{fs}$, and energy of $0.8~\text{mJ}$ is focused into a nitrogen gas cell. 
As the fundamental pulse propagates, it passes through a beta barium borate (BBO) crystal, and a copropagating second-harmonic ($400~\text{nm}$) pulse is generated. 
The fundamental and second-harmonic pulses will be referred to together as the ``pump pulses." 
The pump pulses largely overlap both spatially and temporally as they approach their common focal point. 
When they reach sufficient intensity, they weakly ionize the gas and generate THz radiation.

The THz radiation is generated when the electrons produced by ionization create a cycle-averaged current on the time scale of the pump-pulses' envelope. 
Atoms are preferentially ionized at temporal peaks in the laser field and the resulting electrons are born with essentially zero velocity. 
In a single-color pulse, electrons ionized on either side of the peak field acquire drift velocities in opposite directions. 
The resulting electrons have no ensemble-averaged drift velocity, and therefore no macroscopic, cycle-averaged current.
However, when two colors are present with the appropriate relative phase, they interfere, and electrons acquire a macroscopic, cycle-averaged current.
The cycle-averaged current builds up on the time scale of the pump-pulses' duration and drives the THz fields.
This two-color THz generation mechanism is sometimes couched as a four-wave mixing process, but, strictly speaking, it is not due to a third-order nonlinearity.

There are other mechanisms which can modify the two-color, cycle-averaged current or even produce a cycle-averaged current in the absence of the second color. 
The envelope in few-cycle, single-color laser pulses varies fast enough that a cycle-averaged current on the time scale of the envelope can be created \cite{Xu2013}.  
This current can drive broadband THz radiation similar to the two-color mechanism.  
For laser pulses intense enough to deplete the neutral gas, a cycle-averaged current can be formed. 
This occurs because, during a half cycle, there are more neutral gas molecules to ionize on the rise to the peak field than on the decent. 
The optimal phase for THz generation in intense, two-color pulses can be modified by this effect \cite{Dai2011}.
Both effects are included in our model, but are not significant for the parameters we consider. 
A third effect related to the time variation of the envelope of an elliptically polarized laser pulse is not included in our study, which focuses on linearly polarized fields.

We observe in simulations few-cycle THz pulses that propagate at an angle, $\phi\approx 1^\circ$, above and below the optical axis. 
This can be explained with an optical Cherenkov model, where the cycle-averaged current, created by the pump pulses, moves faster than the THz propagation velocity. 
Optical Cherenkov is a common mechanism for generating THz radiation in electro-optic crystals by the nonlinear optics community \cite{Auston1984}. 
We will also discuss a unification of our Cherenkov model with the ``oscillating current" model introduced by You \textit{et al}.\ \cite{You2012}.  
In this way both effects can be seen as different limits of one model. 
D'Amico \textit{et al}.\ \cite{Amico2008} observed conical THz and it was interpreted as a transition-Cherenkov effect, i.e., a single-color optical pulse drives a collisional-damped, few-cycle plasma oscillation via the pondermotive force. 
The plasma wake following the drive laser emits THz radiation as if it were a dipole aligned with the optical axis, traveling at the speed of the optical pulse.     
This differs from our mechanism in two ways: The cycle-averaged current is transverse to the direction of propagation and is not driven by the pondermotive force.

The organization of this paper is as follows: First we will describe the components of our propagation and material response models. 
During this we will discuss the necessity of including each physical phenomena in our model for studying THz generation. 
Finally, we will describe the Cherenkov model, its connection to the oscillating current model,  and analyze our simulation results.

\section{Model}

\subsection{Unidirectional pulse propagation model}

The optical and THz pulses of interest propagate predominately in the forward direction  \cite{Kohler2011}, justifying the use of the unidirectional pulse propagation equation (UPPE) \cite{Kolesik2004}, where the main assumption is that the backward propagating fields do not contribute to the nonlinear response of the medium. The UPPE is amenable to pseudospectral methods which reduce the electromagnetic propagation equation to a set of coupled ordinary differential equations for the field's spectral components. Since the fields are propagated in the spectral domain, the UPPE captures linear dispersion to all orders, allowing treatment of broadband, multicolor pulses.

The electric field's spectral components $\widehat{E} = \widehat{E}(k_x,z,\omega)$ are propagated along $z$ according to
\begin{equation} \label{equ: UPPE w generic source}
\partial_z \widehat{E} = - i \left [ k_z - \frac{\omega}{v_w} \right] \widehat{E} + \frac{ \widehat{S} }{- 2 i  k_z},
\end{equation}
where
\begin{equation} \label{equ: generic source}
\widehat{S}(k_x,z,\omega) =  - \mu_0 \omega^2\widehat{P}^\text{(NL,gas)} +  \mu_0\widehat{\pd{J}{\tau}} + i \mu_0 \omega \widehat{J_\text{loss}}.
\end{equation}
 The variables $\omega$ and $k_x$ are Fourier conjugates to the time coordinate in a window moving with velocity $v_w$, $\tau = t-z/v_w$, and the transverse dimension, $x$, respectively. The medium's nonlinear response to the field, $S(x,z,\tau)$, is calculated in the $(x,\tau)$ domain and then transformed to the spectral domain, $\widehat{S}=\widehat{S}(k_x,z,\omega)$, to drive the fields.
The $z$ component of the wave number, $k_z=k_z(k_x,\omega)$, depends on the frequency and  transverse wave number, and includes the linear response of the gas through the refractive index, $n(\omega)$. 
Specifically, $k_z(k_x,\omega) = \sqrt{\omega^2 n(\omega)^2/c^2 - k_x^2}$. 
The propagation constant in Eq.\ \eqref{equ: UPPE w generic source}, $k_z-\omega/v_w$, reflects the shift in the $z$ component of the wave number due to the moving window.
The nonlinear response of the medium can be decomposed into a bound nonlinear response of the neutral gas $\widehat{P}^\text{(NL,gas)}$, the free electron response $\widehat{\partial_\tau J}$, and an effective current to deplete the field energy during ionization, $\widehat{J_\text{loss}}$.

\subsection{Material Response of Molecular Nitrogen}

The frequency dependent refractive index for molecular nitrogen, $n(\omega) = 1 + \delta n_\text{PK}(\omega)$, in the range $106 - 549 \THz$ ($2.8 - 0.5 \micron$) is given by an equation fit to experimental data and is provided by Peck and Khanna \cite{Peck1966},
\begin{equation} \label{equ: Peck Khanna}
10^8 \delta n_\text{PK}(\omega) = 6497.378 + \frac{ 3073864.9~\micron^{-2}}{144~\micron^{-2}  - \left(\omega/ 2 \pi c\right)^2}.
\end{equation}
For frequencies below $106 \THz$, the index is found by extrapolating Eq.\ \eqref{equ: Peck Khanna}. Recent experiments in air \cite{Lu2012} have indicated $n_\text{air}-1 \approx 1.7 \times 10^{-4}$ at THz frequencies, which is similar to the zero frequency limit of Eq.\ \eqref{equ: Peck Khanna}, $n(0)-1 = 2.78 \times 10^{-4}$. By extrapolating Eq.\ \eqref{equ: Peck Khanna}, the detailed structure in the refractive index due to vibrational and rotational excitations of {\nitrogen} is not included.

The nonlinear bound response of neutral {\nitrogen} is captured in the nonlinear polarization density, $\widehat{P}^\text{(NL,gas)}$, and is calculated in the $(x,\tau)$ domain using
\begin{equation}\label{equ: nonlinear polarization}
P^\text{(NL,gas)} = \frac{4}{3} c  \epsilon_0^2 n_2^\text{(inst)} E^3 + \epsilon_0 n_0 \Delta\alpha Q E.
\end{equation}
Here, two third-order nonlinear processes contribute to the polarization density: an instantaneous electronic response and a delayed rotational response, the first and second terms of Eq.\ \eqref{equ: nonlinear polarization}, respectively. 
In a classical picture of the instantaneous nonlinear bound response, the laser field strongly drives bound electrons and they experience the anharmonicity of the binding potential. 
Because gases are isotropic on macroscopic scales, the lowest-order nonlinear polarization to manifest itself at macroscopic scales is proportional to $E^3$, instead of $E^2$. 
We use $n_2^\text{(inst)} = 7.4 \times 10^{-20}~\text{cm}^2\text{/W}$ at a {\nitrogen} density of $n_0 = 2.5 \times 10^{19} ~\unit{cm}^{-3} $ \cite{Wahlstrand2012}. 
The delayed response arises because the laser field applies a torque to the {\nitrogen} molecules due to the anisotropy in their linear polarizability, $\Delta\alpha = \alpha_\parallel - \alpha_\perp = 6.7 \times 10^{-25}~\text{cm}^3$, where $\alpha_{\parallel,\perp}$ are the linear polarizabilities parallel and perpendicular to the molecular axis, respectively.  
A simple model for the molecular alignment of the gas, $Q=Q(x,z,\tau)$, is to treat it as a driven, damped, harmonic oscillator:
\begin{equation} \label{equ: molecular alignment}
\pdd{Q}{\tau} + 2 \nu \pd{Q}{\tau} + \Omega^2 Q = 2 \Omega^2 n_2^\text{(align)} \epsilon_0 c E(\tau)^2.
\end{equation}
The oscillator parameters $\nu = 9.6 \THz$, $\Omega=18 \THz$, $n_2^\text{(align)}=1.35\times10^{-15}~\text{cm}^2\text{/W}$ are chosen to best match density matrix calculations \cite{Palastro2012} where the laser pulse duration, $\approx25~\text{fs}$, is much shorter than the thermal rotational time scale, $2\pi/\Omega$. 
These two nonlinear processes result in propagation effects such as spectral broadening, harmonic generation, and self-focusing.

During propagation of high power, ultrashort laser pulses, field ionization is the primary mechanism for free electron generation. 
This can be modeled with a rate equation for the electron density, $n_e=n_e(x,z,\tau)$, where 
\begin{equation}
\pd{n_e}{\tau} = w \left( n_0 - n_e\right).
\end{equation}
The rate of electron generation is the ionization rate of a single molecule, $w=w[E(x,z,\tau)]$, times the number density of neutral molecules, $n_n=n_0-n_e$, where $n_0$ is the initial density of the neutral gas.
Here we neglect electron transport, recombination, and attachment; the time scales for these processes are much longer than the pump-pulses' duration \cite{Sprangle2004}.

We use a two-color hybrid ionization rate, $w[E]$, which is a fit to a Perelomov, Popov, and Terent'ev (PPT) ionization rate \cite{Popruzhenko2008} when $w[E]$ is cycle averaged.
The ionization rate includes multiphoton ionization (MPI) for the two pump-pulse frequencies and tunneling ionization (TI). 
MPI is an $N$th-order perturbative process in the intensity, where a bound electron escapes from its binding potential by absorbing $N$ photons with energy $\hbar \omega$ and frequency $\omega$. 
The energy in the $N$ photons must be greater than or equal to the binding energy $U_i$; $N \hbar \omega \geq U_{i}$. 
Tunneling ionization occurs when the instantaneous electric field deforms the binding potential enough to create a classically allowed region outside the atomic or molecular core. 
With some probability, an electron can tunnel through the barrier between the classically bound and classically free regions, resulting in a free electron.
Further details of the two-color hybrid rate and how it was fit to the limiting cases are given in the Appendix.

The free electron current $J=J(x,z,\tau)$ is determined by the electron momentum balance equation, 
\begin{equation} \label{equ: plasma current}
\pd{J}{\tau} = \frac{e^2}{m_e} n_e E - \nu_{en} J.
\end{equation}
It is through this current that the THz will be generated. 
In Eq.\ \eqref{equ: plasma current}, the electron density is time dependent due to ionization.
There is no momentum source term accompanying the ionization because we assume that new free electrons are born at rest.
It can be shown that the solution of this equation for the macroscopic current is equivalent to the single particle picture of Kim \textit{et al}.\ \cite{Kim2009, Babushkin2010}. 
We include a fixed collision frequency, $\nu_{ne} = 5\THz$, to account for electron-neutral collisions which dominate electron-ion collisions in a weakly ionized gas. 
The collision frequency of $5\THz$ is found by approximating the neutral {\nitrogen} density as atmospheric density and assuming that the electron's temperature is approximately the quiver energy at field intensities of $10^{13} - 10^{14}\WOverCmSqr$ \cite{Sprangle2004}.

The second source term for the electromagnetic fields [see Eq.\ \eqref{equ: generic source}] is the Fourier transform of the time derivative of the current, $\partial_\tau J$.
Care must be exercised in its numerical evaluation. 
If $J$ is solved for in the time domain and then Fourier transformed, the moving window must extend several collision times, $\nu_{ne}^{-1}$, so that the currents decay to zero. 
If the domain is too short, the current is finite at the window boundary and its frequency spectrum has an unphysical  $\omega^{-1}$ dependence. 
To circumvent this, we Fourier transform $n_e E$, which tends to zero outside of the temporal range of the pump pulses', and compute the Fourier transform of $\partial_\tau J$ via
\begin{equation} \label{equ: plasma current in frequency domain}
 \widehat{\pd{J}{\tau}} =  \frac{e^2}{m_e}  \frac{\widehat{n_e E}}{1 - i \nu_{en}/\omega}.
\end{equation}

During ionization, the electric field must perform work equal to the ionization potential $U_i$ to liberate each electron.
Ionization energy depletion is included by adding an effective current, $J_\text{loss}=J_\text{loss}(x,z,\tau)$, that accounts for the rate of energy loss: $E J_\text{loss} = w[E] n_n U_i$  \cite{Sprangle2002} ,
\begin{equation}
J_\text{loss} = \frac{w[E] n_n U_i}{E}.
\end{equation}
To avoid issues when dividing the cycle-averaged contributions of Eq.\ \eqref{equ: hybrid ionization model} by the instantaneous electric field, the loss current is only evaluated when $\abs{E(t)} > 27~\text{MV/cm}$. Below these field strengths, the ionization rate is too small to significantly deplete the pump pulses.

\section{Results}
We now describe simulation results based on the numerical solution of the model equations introduced in the previous section. 
The incident electric field is composed of two pulses with central wavelengths $\lambda=800$ and $400~\text{nm}$, respectively. 
The $800~\text{nm}$ pulse has a total energy of $0.7~\text{mJ}$, a full-width half-maximum duration of $25~\text{fs}$, and a vacuum spot size of $w_0 = 15.3 \micron$.  
The $400~\text{nm}$ pulse is created experimentally by second-harmonic generation in a BBO crystal. 
This motivates the $400~\text{nm}$ pulse having a total energy that is 10\% of the fundamental pulse, $0.07~\text{mJ}$, a full-width half-maximum duration that is a factor $\sqrt{2}$ shorter than the fundamental, $18~\text{fs}$,  and a vacuum spot size that is $\sqrt{2}$ smaller than the fundamental, $w_0 = 11 \micron$. 
The pulses are assumed to overlap spatially and temporally with the peak of each pulse colocated $8~\text{cm}$ before the vacuum focus.  
This is where the BBO crystal ends and the simulation begins. 
Both colors are initialized with a phase front curvature that is consistent with passing through a lens with focal length and diameter of $15$ and $0.5~\text{cm}$, respectively. 
The polarization of the pump pulses are assumed to be collinear.

The simulation domain is $6~\text{mm}$ in the transverse spatial dimension, $x$, and $1~\text{ps}$ in the time domain, $\tau$, with $2^9$ and $2^{15}$ grid points, respectively. 
The transverse spatial resolution is $\Delta x =12 \micron$.
This resolution is sufficient because plasma refraction keeps the pulse from reaching its vacuum spot size. 
For example, the pump-pulses' time-averaged rms radii is always larger than $100 \micron$.
At the front of the pulse, where the intensity is lower, the rms radius reaches a minimum of   $40 \micron$.
The transverse spatial resolutions also resolve the transverse phase variation associated with focusing sufficiently well for the vacuum focal point to remain unchanged.
Simulations with double the spatial resolution, $\Delta x=6 \micron$, show convergence of the THz energy and fields. 
The temporal domain is chosen so as to capture low frequency behavior, $\Delta f = 1~\text{THz}$, while having sufficiently small time steps, $\Delta \tau = 0.03~\text{fs}$, to resolve ionization bursts and harmonic generation. 
The pulses propagate $12~\text{cm}$, with a uniform step size of $\Delta z  = 10 \micron$. 
The window velocity, $v_w=0.99972 c$, is comoving with the group velocity of $800~\text{nm}$ in {\nitrogen}. 
The background {\nitrogen} density is $n_\text{gas}=2.5\times 10^{19}~\text{cm}^{-3}$. 
The UPPE model, Eq.\ \eqref{equ: UPPE w generic source}, is solved using a second-order predictor-corrector scheme for the nonlinear term, $\widehat{S}$.

\begin{figure}[ht]
	\begin{center}
	\includegraphics[width=3.2in,draft=false]{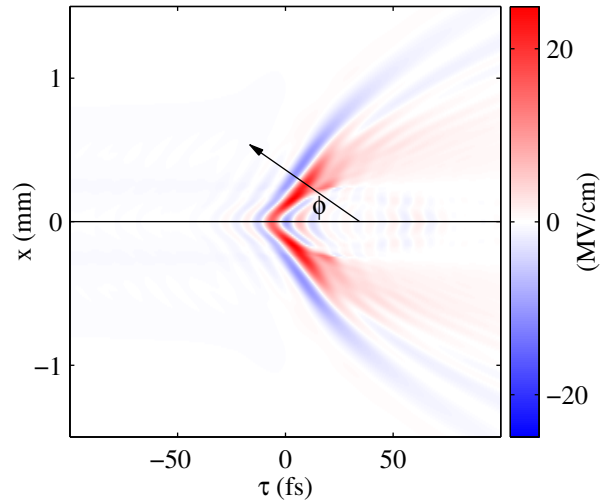}
    \caption{The electric field from 0 to $100\THz$ is shown in the transverse spatial dimension, $x$, versus a time window that is comoving with the $800~\text{nm}$ pulse, $\tau$. The pulse is propagating from right to left with an off-axis angle $\phi$. The electric field at $2~\text{cm}$ before vacuum focus was chosen because most of final THz energy is already in the pulse.  \label{fig: Conical THz Example}}
    \end{center}
\end{figure}

The simulation predicts off-axis, broadband, THz radiation as seen in Fig.\ \ref{fig: Conical THz Example}. 
The figure displays the THz electric field as a function of $x$ and $\tau$ after propagating to $2~\text{cm}$ before the vacuum focus.  
To calculate the THz electric field, $\widehat{E}$ has been filtered to remove frequency components with $f>100\THz$ and transformed to the space and time domain.
The THz field is a few-cycle pulse that has been created near the axis and is propagating at approximately 1$^\circ$ above and below the propagation axis of the pump pulses. 
This can be seen from the nulls in the phase (white in the figure) where the fields will propagate perpendicular to the phase front. 

\subsection{Cherenkov Model}

\begin{figure}
	\centering
    \includegraphics[height=3.2in,draft=false]{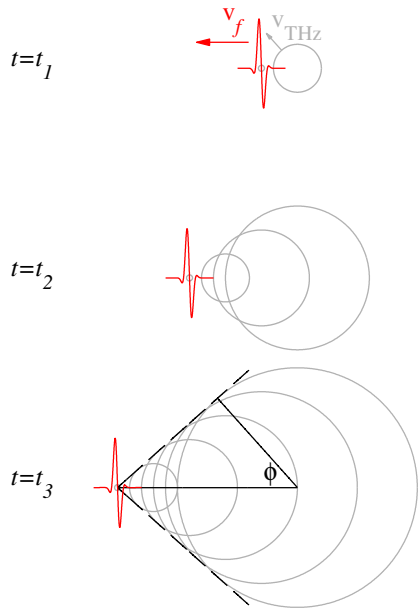}
    \caption{The broadband THz frequency current (red) is traveling faster from right to left than the phase velocity of the THz fields (lines of constant phase are shown in gray). Constructive interference can be seen along the front (black dashed) above and below the propagation axis.
    \label{fig: Cherenkov Cartoon}}    
\end{figure}

The angle of the THz pulse shown in Fig.\ \ref{fig: Conical THz Example} can be explained by an optical Cherenkov effect.
As the pump pulses approach focus, their fronts of constant intensity and, through ionization, fronts of constant plasma density move axially faster than the pump-pulses' group velocities. 
The resulting current drives the THz radiation and travels faster than the THz phase velocity in the medium. 
This results in a ``Cherenkov cone" in which the emitted THz field interferes constructively at the Cherenkov angle $\phi$ given by $\cos{\phi} = v_{\text{THz}}(\omega)  / v_f$, where $v_f$ is the velocity of the plasma current front and $v_{\text{THz}}(\omega) = c / n(\omega)$ is the THz phase velocity. 
A schematic of this is shown in Fig.\ \ref{fig: Cherenkov Cartoon}.
The duration of the current approximates the time scale of the pump-pulses' envelope, providing the few-cycle THz phase front observed in Fig.\ \ref{fig: Conical THz Example}.

A simple model illustrates this phenomenon.
Equations \eqref{equ: UPPE w generic source} and \eqref{equ: generic source} can be solved analytically to find the THz field spectrum resulting from a prespecified THz current.
We model the current driven by the pump pulses as a localized, on-axis source with velocity,  $v_f$,  $J_{p}(x,z,t) = I_0 \delta (x) \theta (t - z/v_f)$ where $I_0$ is the current amplitude ($\text{A m}^{-1}$ in two dimensions). 
After the current pulse has propagated a distance, $L$, the THz spectrum is given by
\begin{equation}
\left | \widehat{E}_{\text{THz}} (k_x,z,\omega)\right |^2 = \frac{I_0^2 \mu_0^2}{16 k_z^2} \sinc^2 \left[ \left( \frac{\omega}{v_f} - k_z \right) L \right] L^2,
\end{equation}
where $k_z = \sqrt{(\omega n / c)^2 - k_x^2}$. 
The peaks in the power spectrum occur approximately where the argument of the $\sinc$ is zero, reproducing the expression for the Cherenkov angle:
\begin{equation}\label{equ: Cherenkov Model}
\cos \phi = v_{\text{THz}}(\omega)  / v_f.
\end{equation}
We note that the THz angle is related to the vector components of the wave number via $k_z = (\omega n / c) \cos{\phi}$.

This model can be extended to capture a current source with transverse spatial extent or a current source that oscillates along the propagation distance. 
The latter extension captures the effect on the two-color THz current of phase slippage between the pump pulses due to their phase-velocity difference.  
This phase slippage was considered in a previous model of off-axis THz emission \cite{You2012}. 
You \textit{et al}.\ \cite{You2012} treat the THz driving current as a dipole radiator traveling with the laser pulse. 
The phase of the dipole's oscillation, and hence the emitted radiation, varies along the propagation axis with the relative phase between the pump pulses. 
In You's model, the group velocity of the laser pulses, the velocity of the driving current, $v_f$, and the THz phase velocity, $v_\text{THz}$, are all set to $c$. 
While the model predicts off-axis radiation, the equality of THz and drive velocities precludes Cherenkov radiation. 
Our model can capture this oscillating current effect if we impose a second spatial variation on the current density,  $J_{p}(x,z,t) = I_0 \delta (x) \cos (k_d z) \theta (t - z/v_f)$.
In this case the THz spectrum is peaked at angles given by
\begin{equation}\label{equ: General Off Axis Model}
\cos \phi  =  v_\text{THz} / v_f \pm  k_d v_\text{THz} / \omega,
\end{equation}
where $\omega$ is the THz frequency of interest, $k_d=\pi/L_\pi$ is the dephasing wavenumber, and $L_\pi$ is the distance over which the two colors will phase slip by $\pi$.

The dephasing length is inversely proportional to the phase-velocity difference and can be estimated as $L_\pi = (\lambda_0 /4 ) \abs{n(\omega_0)-n(2\omega_0)}^{-1}$ \cite{Rodriguez2010}. 
The refractive index is given by $n(\omega) = 1 + \delta n_\text{gas}(\omega) + \delta n_\text{plasma}(\omega) + \cdots$, where $\omega$ could be for either the fundamental, $\omega_0$, or second harmonic, $2\omega_0$. 
The quantity $\lambda_0$ is the wavelength of the fundamental. 
From {\nitrogen} dispersion alone $L_\pi = 2.7~\text{cm}$, but with a plasma density in the range of $10^{16} - 10^{17}~\text{cm}^{-3}$, the dephasing length would be $2.1 - 0.7~\text{cm}$, respectively. These plasma densities are typical for the region where THz is generated.

\begin{figure}
	\centering
    \includegraphics[width=3.2in,draft=false]{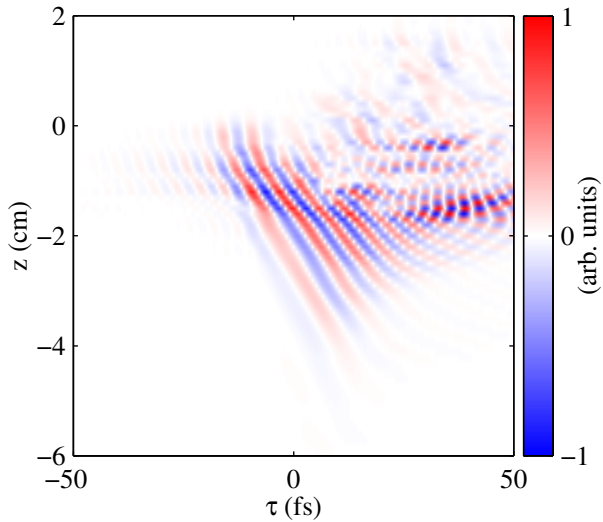}
    \caption{The on-axis $\partial_\tau J$ after a low-pass filter with cutoff frequency of $200\THz$ has been applied. 
    \label{fig: THz Source} }
\end{figure}

\begin{figure}
	\centering
    \includegraphics[width=3.2in,draft=false]{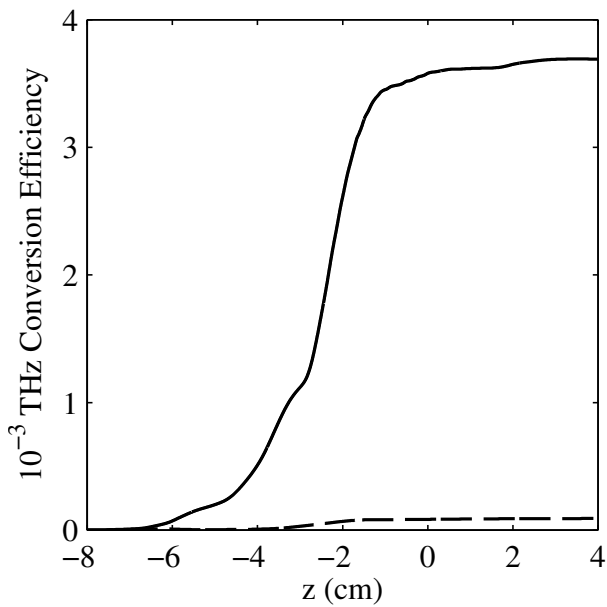}
    \caption{The solid line is the energy in the THz from $10$ to $100\THz$ relative to the total initial energy in the pump pulses. The dashed line is for the same simulation parameters, but only the nonlinear gas response was allowed to drive THz radiation. \label{fig: THz Energy}}
\end{figure}

Figure \ref{fig: THz Source} displays the time derivative of the current density on axis, low-pass filtered to frequencies below $200\THz$ as a function of $z$ and $\tau$. 
Most of the THz energy is generated between $z=-4$ and $-1~\text{cm}$, as can be seen in Fig.\ \ref{fig: THz Energy}. 
Over this distance, the THz current source has the form of a temporally oscillating signal that moves forward in the frame of the simulation. 
For comparison, an object moving at the group velocity of $800~\text{nm}$ would trace out a vertical path in $(z,\tau)$ domain, while objects moving faster, or slower, follow paths to the left, or right, of vertical respectively. It is the overall forward motion of the THz $\partial_\tau J$ that drives the Cherenkov radiation.  
The forward motion of the THz current density profile can be attributed to the fact that the pump pulses are converging towards focus. 
As the pulses converge, their intensity rises, and the time in the pulse envelope when ionization becomes significant moves forward in the plane of Fig.\ \ref{fig: THz Source}.

The spatiotemporal form of the current density waveform implied by Fig.\ \ref{fig: THz Source} is that of a few-cycle pulse. 
The temporal $(\approx 10~\text{fs})$ variations in $\partial_\tau J$ at fixed $z$ are due to a combination of the temporal variation in the pump-pulses' relative phase during the pulse and the frequency upshift of the THz field due to the rising electron density. 
We note the variations become more rapid with propagation distance.
As the pump pulses propagate their relative phase becomes a time varying function due to the rise in electron density during the pulses.
The sign of the two-color driven THz current then varies with this relative phase.
This variation becomes more rapid with propagation distance. 
A second contribution to the increase in frequency of the on-axis $\partial_\tau J$ as a function of propagation distance is the direct spectral blueshifting (up to approximately $150\THz$) of the THz fields in the region of increasing free electron density.

The signal in Fig.\ \ref{fig: THz Source} was low-pass filtered at $200\THz$ (as opposed to the $100\THz$ filter applied in Fig. \ref{fig: Conical THz Example}) to include the peak frequency of the on-axis, blue-shifted THz field (around $150\THz$ at $z=-2~\text{cm}$). 
While the peak frequency is larger on axis, most of the THz field energy is distributed off axis where the average frequency is lower ($\approx 50 \THz$).

The front velocity is extracted from Fig.\ \ref{fig: THz Source} by measuring the slope of the null lines of $\partial_\tau J$. 
We find that the front velocity is approximately $v_f=0.999\,95c$.
For comparison, the $800~\text{nm}$ group velocity is $v_{g,800~\text{nm}} = 0.999\,72c$. 
With this front velocity and the refractive index model discussed above, the Cherenkov model predicts an off-axis angle of $\phi \approx 1.2^\circ$ [according to Eq.\ \eqref{equ: Cherenkov Model}], which is similar to $0.9^\circ$, the value seen in Fig.\ \ref{fig: Conical THz Example}. 

Finally, we note the space-time dependence of the time derivative of the current density is not of the form required to produce Eq. \eqref{equ: General Off Axis Model} (except when $k_d\approx0$).
There is variation of the waveform with $z$, in addition to translation at $v_f$.
The amplitude of $\partial_\tau J$ grows and the frequency increases over a distance of $3~\text{cm}$. 
However the behavior is not a periodic oscillation with a clearly identifiable wave number $k_d$.

\subsection{Angular dependence of THz on refractive index}

To test the model giving rise to Eq.\ \eqref{equ: Cherenkov Model} we attempt to vary $v_\text{THz}$. 
Competing propagation effects in the simulation make control of the current front velocity challenging. 
The THz phase velocity, on the other hand, can be directly manipulated by modifying the refractive index at THz frequencies. 
The resulting change in the simulated THz emission angle can then be compared to predictions of the Cherenkov model. 
Specifically, we use the following modified refractive index model;
\begin{equation} \label{equ: ref index for modified low freq}
\delta n(\omega) = 
\begin{cases}
	 \delta n_{0},   		&	\omega/2\pi<  190\unit{THz} \\
     \delta n_\text{PK}(\omega),	&	\text{otherwise},
\end{cases}
\end{equation}
where $n(\omega) = 1 + \delta n(\omega)$ and $\delta_\text{PK}(\omega)$ is defined in Eq.\ \eqref{equ: Peck Khanna}. 
While the modified refractive index has no frequency dependence below $190\THz$, the relative change in the actual refractive index of {\nitrogen} is only $0.2\,\%$ between $0$ and $190\THz$ \cite{Peck1966}. 
In all cases, the group velocity at low frequencies in {\nitrogen} is not significantly different than the phase velocity. 
Experimentally, the dispersion at low frequencies could be modified by the selection and relative percentage of gas species in the medium.

\begin{figure*}[!ht]
                \includegraphics[width=6.4in,draft=false]{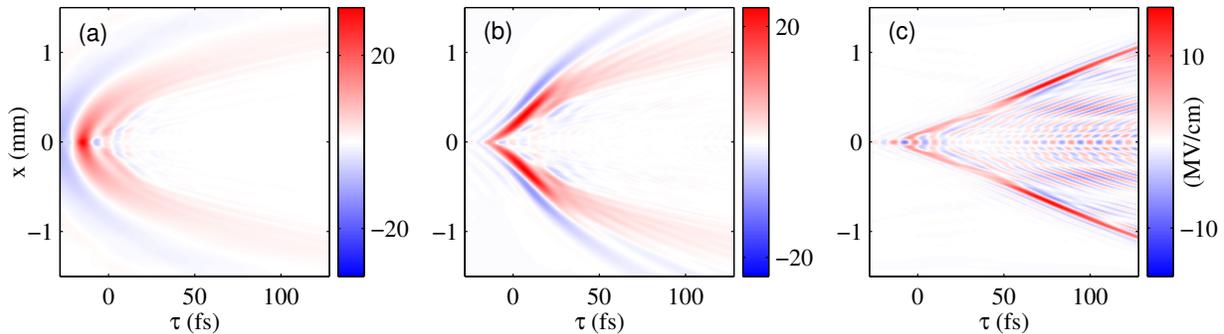}
        \caption{Shows the electric field at $1\,\text{cm}$ before vacuum focus from $0$ to $100\THz$ for three different THz dispersion models; $\delta n_{0}$ = $0$, $2.78 \times 10^{-4}$, and $1.1 \times 10^{-3}$ for (a), (b), and (c) respectively. Most of the THz have been generated at this point. 
        \label{fig: THz Fields vs index}}
\end{figure*}

Figure \ref{fig: THz Fields vs index} shows the extracted THz electric field for $\delta n_{0}$ = 0, $2.78 \times 10^{-4}$, and $1.1 \times 10^{-3}$. 
The propagation angle of the THz radiation can be seen to increase with increasing $\delta n_{0}$, as anticipated by Eq.\ \eqref{equ: Cherenkov Model}. 
The variations of $\delta n_0$ leave the pump pulses and current front velocity largely unchanged.
The pump pulses drive the current source and indirectly control the front velocity. 
Changes to the pump-pulses' propagation, due to changes in the THz refractive index, should only occur via nonlinear interactions with the THz frequencies, e.g.\ non-degenerate four-wave mixing. 
These interactions tend to be smaller than the nonlinear processes involving the pump pulses alone.

\begin{figure}
	\centering
    \includegraphics[width=3.2in,draft=false]{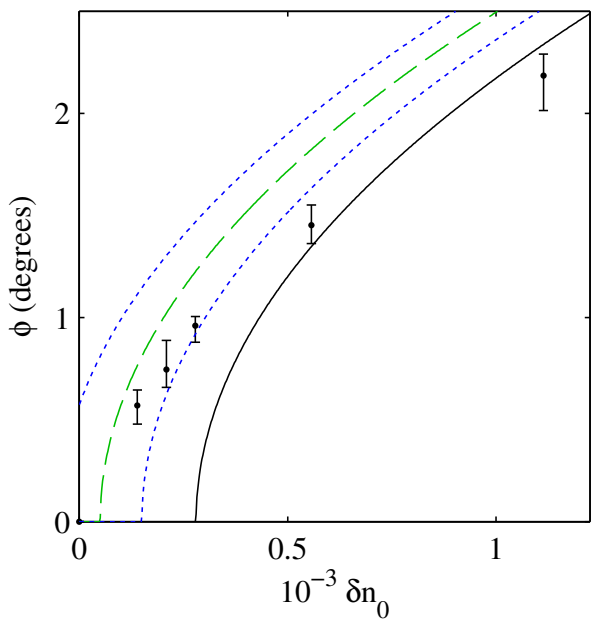}
    \caption{
    The dots with error bars are measured THz angle at $z=-1~\unit{cm}$ for separate simulations with a refractive index given by Eq.\ \eqref{equ: ref index for modified low freq}. 
    The curves are Eq.\ \eqref{equ: Cherenkov Model} with fixed $v_f$ and $v_\text{THz}$ determined by the refractive index at $\omega=0$.  
    The solid black and dashed green curves are specifically for $v_f =0.999\,72 c$ and $v_f = 0.999\,95 c$.
    The dotted blue curves are from Eq.\ \eqref{equ: General Off Axis Model} when the frequency is $50\THz$, the dephasing length is $3~\text{cm}$, and the front velocity is $0.999\,95 c$.
    \label{fig: THz Angle}}    
\end{figure}

The dependence of the THz propagation angle, $\phi$, on $\delta n_{0}$ is shown in Fig.\ \ref{fig: THz Angle}.
For each $\delta n_{0}$, the THz angle is extracted from images such as those in Fig.\
 \ref{fig: THz Fields vs index} after most of the THz radiation has been generated, $z=-1~\unit{cm}$. 
The simulation results are bounded by the Cherenkov model, Eq.\ \eqref{equ: Cherenkov Model}, evaluated with $v_f$ equal to the group velocity of $800~\unit{nm}$ ($0.999\,72c$) and the extracted front velocity, $v_f=0.999\,95$, from the simulations. 
This shows reasonable agreement between the predicted Cherenkov model and our simulations.
The blue dotted curves in Fig.\ \ref{fig: THz Angle} show the predicted angle for the positive (lower curve) and negative (upper curve) solutions of Eq.\ \eqref{equ: General Off Axis Model}.
We substitute $k_d=\pi/L_\pi$ with $L_\pi=3~\text{cm}$ which is roughly the distance over which the THz current waveform varies. 
In this way Eq.\ \eqref{equ: General Off Axis Model} can be used to indicate the degree of uncertainty in the prediction of Eq.\ \eqref{equ: Cherenkov Model}.

\subsection{Cherenkov radiation from four-wave mixing}

\begin{figure}
	\centering
    \includegraphics[width=3.2in,draft=false]{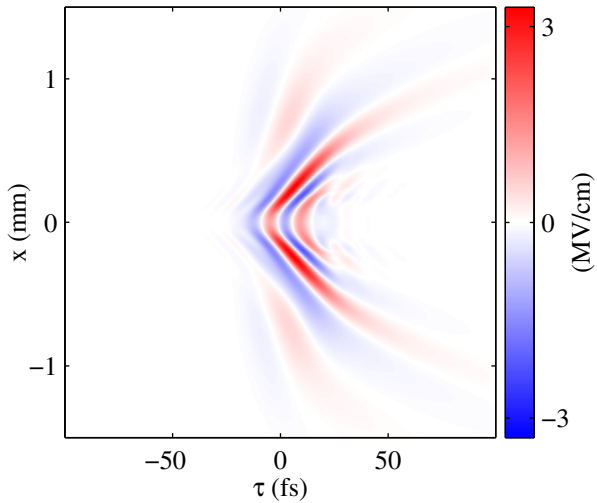}
    \caption{The electric field from 0 to $100 \THz$ is shown in the transverse spatial dimension, $x$, versus a time window that is co-moving with the $800~\text{nm}$ pulse, $\tau$. This THz electric field comparable to that of Fig.\ \ref{fig: Conical THz Example}, except that this one was generated exclusively by a four-wave rectification process.  \label{fig: Conical THz from FWM}}    
\end{figure}

In simulations, the free electron current is the dominant mechanism for generation of THz radiation \cite{Berge2013}.
When the current source, $\partial_\tau J$, and the effective loss current are removed from Eq.\ \eqref{equ: generic source} using a high-pass filter, the third-order nonlinearities [the first term in Eq.\ \eqref{equ: generic source}] still generate THz radiation as seen by the dashed curve in Fig.\ \ref{fig: THz Energy}. 
But in this scenario, the conversion efficiency from pump-pulses' energy to THz is a factor of $\approx 40$ times smaller than the photocurrent model. This is similar to results reported in \cite{Berge2013}.
Interestingly, the THz generated via four-wave mixing is also conical, suggesting that the optical Cherenkov mechanism is still at play. 
Figure \ref{fig: Conical THz from FWM} shows the THz field that is generated from four-wave interaction alone. The THz angle is the same as that of Fig.\ \ref{fig: Conical THz Example}.
This is expected since the bound nonlinear polarization current, which drives the THz, will follow the superluminal intensity fronts of the pump pulses. 

\subsection{Experimental comparison}

While the simulations seem to predict a THz propagation angle of $\approx 1^\circ$, You \textit{et al}.\ observe THz radiation at angles of $\approx 4^\circ$ \cite{You2012}. 
In the experiment, the focus was on frequencies below $10~\unit{THz}$ as opposed to the broadband radiation below $100 \THz$ that we have investigated. 
Blank \textit{et al}.\ \cite{Blank2013} observed a THz intensity spectrum that extends up to $100\THz$ with an off-axis angle of $3.2^\circ$. 
Their experiments are performed in air with similar parameters to ours: a pump-pulse energy of $0.42~\text{mJ}$, fundamental wavelength of $775~\text{nm}$, pump-pulse duration below $20~\text{fs}$, and a focal length of $20~\text{cm}$.   
We find, if we further filter the THz signal, the average off-axis angle from the electric field power spectrum for frequencies between $5$ and $10\THz$ to be $2.1^\circ\pm1.0^\circ$.
This is closer to the experimentally measured values. 
Differences still remain between the conditions in our simulations and the experiments.
The simulated medium is {\nitrogen} as opposed to air. 
The index of refraction of air in the $10\THz$ range may have a frequency dependence not contained in our simulations. 
Also, the presence of oxygen, with a lower ionization potential than {\nitrogen}, could lead to more free electrons and a different THz current source speed. 
Finally, the simulations are two dimensional.
The superluminal front velocity is due to the focusing of the pump pulses.
This speed can then be altered in going from two to three dimensions.

\section{Conclusion}

We have developed a two-dimensional, unidirectional, electromagnetic propagation code to examine two-color THz generation in {\nitrogen}.
The model includes linear dispersion to all orders, the instantaneous and delayed-rotational nonlinear bound response, free electron generation via multiphoton and tunneling ionization, plasma response including collisional momentum damping, and ionization energy depletion.
We have found that the off-axis, THz generation predicted by the simulations can be explained as an optical Cherenkov process. 
The angle of THz emission depends sensitively on the low frequency refractive index and current front velocity. 
Using our best estimate of the frequency dependent refractive index produces reasonable agreement with the experiment. 
Although the THz radiation is generated predominately by the photocurrent mechanism, the Cherenkov process also determines the emission angle of THz radiation generated by two-color, four-wave interaction in the nonlinear molecular polarizability.

\appendix
\section*{Appendix: Hybrid Ionization Rate}\label{sec: ionization model}

MPI and TI are distinct limiting cases of a more general nonlinear photoionization theory such as that of Keldysh \cite{Keldysh1965, Popov2004} or later refinements by PPT and others \cite{Perelomov1966, Popruzhenko2008}.
These limiting cases are roughly delineated by the Keldysh parameter $\gamma=\omega\sqrt{2 m_e U_i}/e \mathcal{E}$, where $m_e$ and $e$ are the electron mass and charge, while $\mathcal{E}$ is the electric field amplitude.
For example, $\gamma \gg 1$ implies the multiphoton regime, while $\gamma \ll 1$ implies the tunneling regime.

As the pump pulses focus, the field strength will transition from the multiphoton to the tunneling regime.
In the multiphoton regime, the TI rate underestimates free electron generation.
Therefore, the decreased refractive index associated with the multiphoton generated free electrons can defocus the pump pulses and modify subsequent propagation more than expected from a TI rate. 
Unfortunately, the PPT ionization rate, which covers both regimes, is for a single color and dependent on the intensity, not on the instantaneous electric field.
Therefore, it does not generate THz radiation according to the mechanism of interest.

The motivation for the hybrid ionization rate is to capture both the instantaneous nature of the tunneling ionization rate when in the tunneling regime, while not significantly underestimating free electron generation and defocusing effects when in the multiphoton regime.

\begin{figure}
	\centering
    \includegraphics[width=3.2in,draft=false]{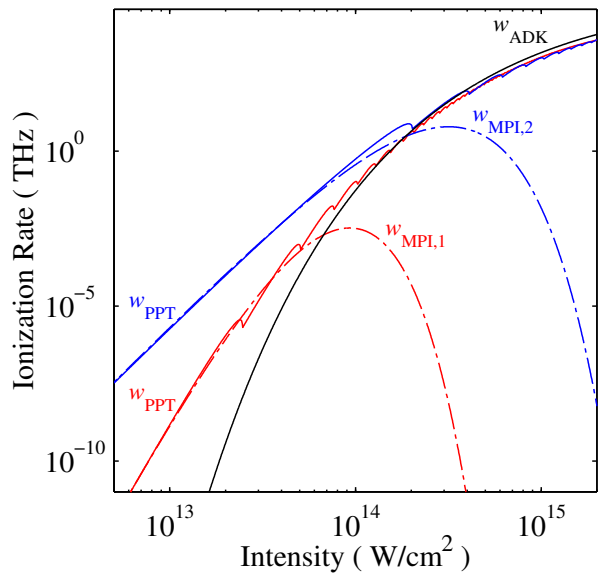}
    \caption{ The solid red and blue curves represent PPT ionization rates for $\lambda=800\,\text{nm}, ~ 400\,\text{nm}$ respectively. The solid black curve indicates a cycle-averaged tunneling rate which approaches the PPT rate at high intensities. The dashed-dotted red and blue curves show the MPI rates for $\lambda=800\,\text{nm}, ~ 400\,\text{nm}$ respectively. Notice that a single color MPI rate plus the tunneling rate is a reasonable approximation of the associated PPT rate.
    \label{fig: ionization rate}}    
\end{figure}

Conventional MPI rates depend on the intensity to a large power \cite{COUAIRON2007}.
This poses a problem when attempting to approximate the PPT ionization rate by interpolating from the multiphoton to the tunneling regime, e.g., by summing the MPI and TI rates. 
The problem arises because the MPI rate is orders of magnitude larger than the TI rate when evaluated in either the multiphoton or tunneling regimes. 
Therefore, the sum of the individual rates is always dominated by MPI.
This is beneficial in the multiphoton limit but not in the tunneling limit where the tunneling rate should be a reasonable approximation.
We adapt the MPI rate to drop exponentially with increasing intensity, as shown by the dashed-dotted red and blue curves of Fig.\ \ref{fig: ionization rate}.
The modified MPI rate is then summed with the tunneling ionization rate to yield our single-color hybrid ionization rate.
The cutoff intensity used in the exponential decay, $I_{\text{cutoff}}$, becomes a free parameter that is used to match $w_{\text{MPI,i}} + w_\text{ADK}$, after cycle averaging, to the PPT ionization rate for each color \cite{Popruzhenko2008}.

We then extend this hybrid ionization rate for two-color pulses. 
In the tunneling limit, the ionization rate should depend on the instantaneous field and therefore the Ammosov-Delone-Krainov (ADK) model should capture the two-color ionization dynamics \cite{Larochelle1998,Talebpour1999}. 
But in the multiphoton regime, the rate is strongly dependent on the frequency. 
In general, a nonlinear process like ionization is not additive in the individual rates.
It is possible that mixed-photon ionization channels, like those involving $N$ $800~\text{nm}$ photons and $M$ $400~\text{nm}$ photons, would have important contributions to the total ionization rate. 
But summing the $800~\text{nm}$ and $400~\text{nm}$ MPI rates provides a better lower bound on the free electron generation in the multiphoton regime than neglecting either or both.  
Additionally, it provides a rate that can be fit to the accepted PPT rates in the limits of a laser pulse of either color. 
The absence of computationally efficient, quantum mechanical, atomic or molecular response models necessitates approximation. 
To this end, we treat the total MPI rate as the sum of the rates for the individual harmonics.

The full two-color hybrid ionization rate is given by
\begin{equation} \label{equ: hybrid ionization model}
w[E] = w_{\text{MPI},1}(I_1) + w_{\text{MPI},2}(I_2) + w_\text{ADK}(E),
\end{equation}
where $I_1$ and $I_2$ are the enveloped intensities of the fundamental- and second-harmonic pulses, respectively. The individual MPI rates are given by $w_{\text{MPI},i} = \sigma_i I_i^{N_i} \exp \left( - I_i / I_{\text{cutoff},i} \right)$, where $\sigma_1 = 4.47 \times 10^{-140}~\unit{cm}^{22} \unit{W}^{-11} \unit{s}^{-1}$ , $N_1=11$, $I_{\text{cutoff},1}=8.46 \times 10^{12}~\unit{W/cm}^2$,  $\sigma_2 = 2.46 \times 10^{-72}~ \unit{cm}^{12} \unit{W}^{-6} \unit{s}^{-1}$ , $N_2=6$, and $I_{\text{cutoff},2}=5.29 \times 10^{13}~\unit{W/cm}^2$. 
The tunneling rate used is outlined in \cite{Larochelle1998} with an ionization potential of $U_i=15.576~\unit{eV}$ and effective Coulomb barrier $Z_\text{eff} = 0.9$ \cite{Talebpour1999}.

In the tunneling regime, Eq.\ \eqref{equ: hybrid ionization model} approximates the instantaneous ADK tunneling rate \cite{Larochelle1998}. 
In the limit of a single color, either $800$ or $400~\unit{nm}$, Eq.\ \eqref{equ: hybrid ionization model} after cycle averaging approaches the PPT rate for that color \cite{Popruzhenko2008}. 
This implies that in the multiphoton limit and in the limit of a single color, Eq.\ \eqref{equ: hybrid ionization model} also matches the MPI rate.

As a result of enveloping, $I_1$ and $I_2$ do not depend on their respective carrier or carrier-envelope phases.
This is consistent with traditional MPI models, which depend on the cycle-averaged field \cite{COUAIRON2007}.
There has been recent theoretical work on the phase dependence of two-color MPI \cite{Kotelnikov2011}, but it does not lend itself to efficient numerical implementation in an electromagnetic propagation code.

\begin{acknowledgments}
We would like to acknowledge T. Rensink, M. Herrera, and P. Sprangle for fruitful discussions. This work was supported by ONR and DOE. 
\end{acknowledgments}

\bibliography{ConicalTHz}

%Merlin.mbs v4.21 2009-07-09.
\begin{thebibliography}{10}%
\makeatletter
\providecommand \@ifxundefined [1]{%
 \ifx #1\undefined \expandafter \@firstoftwo
 \else \expandafter \@secondoftwo
\fi
}%
\providecommand \@ifnum [1]{%
 \ifnum #1\expandafter \@firstoftwo
 \else \expandafter \@secondoftwo
\fi
}%
\providecommand \enquote [1]{``#1''}%
\providecommand \bibnamefont  [1]{#1}%
\providecommand \bibfnamefont [1]{#1}%
\providecommand \citenamefont [1]{#1}%
\providecommand\href[0]{\@sanitize\@href}%
\providecommand\@href[1]{\endgroup\@@startlink{#1}\endgroup\@@href}%
\providecommand\@@href[1]{#1\@@endlink}%
\providecommand \@sanitize [0]{\begingroup\catcode`\&12\catcode`\#12\relax}%
\@ifxundefined \pdfoutput {\@firstoftwo}{%
 \@ifnum{\z@=\pdfoutput}{\@firstoftwo}{\@secondoftwo}%
}{%
 \providecommand\@@startlink[1]{\leavevmode\special{html:<a href="#1">}}%
 \providecommand\@@endlink[0]{\special{html:</a>}}%
}{%
 \providecommand\@@startlink[1]{%
  \leavevmode
  \pdfstartlink
   attr{/Border[0 0 1 ]/H/I/C[0 1 1]}%
   user{/Subtype/Link/A<</Type/Action/S/URI/URI(#1)>>}%
  \relax
 }%
 \providecommand\@@endlink[0]{\pdfendlink}%
}%
\providecommand \url  [0]{\begingroup\@sanitize \@url }%
\providecommand \@url [1]{\endgroup\@href {#1}{\urlprefix}}%
\providecommand \urlprefix [0]{URL }%
\providecommand \Eprint[0]{\href }%
\@ifxundefined \urlstyle {%
  \providecommand \doi [1]{doi:\discretionary{}{}{}#1}%
}{%
  \providecommand \doi [0]{doi:\discretionary{}{}{}\begingroup
  \urlstyle{rm}\Url }%
}%
\providecommand \doibase [0]{http://dx.doi.org/}%
\providecommand \Doi[1]{\href{\doibase#1}}%
\providecommand \bibAnnote [3]{%
  \BibitemShut{#1}%
  \begin{quotation}\noindent
    \textsc{Key:}\ #2\\\textsc{Annotation:}\ #3%
  \end{quotation}%
}%
\providecommand \bibAnnoteFile [2]{%
  \IfFileExists{#2}{\bibAnnote {#1} {#2} {\input{#2}}}{}%
}%
\providecommand \typeout [0]{\immediate \write \m@ne }%
\providecommand \selectlanguage [0]{\@gobble}%
\providecommand \bibinfo [0]{\@secondoftwo}%
\providecommand \bibfield [0]{\@secondoftwo}%
\providecommand \translation [1]{[#1]}%
\providecommand \BibitemOpen[0]{}%
\providecommand \bibitemStop [0]{}%
\providecommand \bibitemNoStop [0]{.\EOS\space}%
\providecommand \EOS [0]{\spacefactor3000\relax}%
\providecommand \BibitemShut [1]{\csname bibitem#1\endcsname}%
%</preamble>
\bibitem{Sherwin2004}%
  \BibitemOpen
  \bibfield{author}{%
  \bibinfo {author} {\bibfnamefont{M.}~\bibnamefont{Sherwin}}, \bibinfo
  {author} {\bibfnamefont{C.}~\bibnamefont{Schmuttenmaer}},\ and\ \bibinfo
  {author} {\bibfnamefont{P.}~\bibnamefont{Bucksbaum}},\ }%
  \enquote{\bibinfo {title} {{Opportunities in THz Science}},}\  (\bibinfo
  {year} {2004})%
  \bibAnnoteFile{NoStop}{Sherwin2004}%
\bibitem{Kampfrath2010}%
  \BibitemOpen
  \bibfield{author}{%
  \bibinfo {author} {\bibfnamefont{T.}~\bibnamefont{Kampfrath}}, \bibinfo
  {author} {\bibfnamefont{A.}~\bibnamefont{Sell}}, \bibinfo {author}
  {\bibfnamefont{G.}~\bibnamefont{Klatt}}, \bibinfo {author}
  {\bibfnamefont{A.}~\bibnamefont{Pashkin}}, \bibinfo {author}
  {\bibfnamefont{S.}~\bibnamefont{M\"{a}hrlein}}, \bibinfo {author}
  {\bibfnamefont{T.}~\bibnamefont{Dekorsy}}, \bibinfo {author}
  {\bibfnamefont{M.}~\bibnamefont{Wolf}}, \bibinfo {author}
  {\bibfnamefont{M.}~\bibnamefont{Fiebig}}, \bibinfo {author}
  {\bibfnamefont{A.}~\bibnamefont{Leitenstorfer}},\ and\ \bibinfo {author}
  {\bibfnamefont{R.}~\bibnamefont{Huber}},\ }%
  \bibfield{journal}{%
  \Doi{10.1038/nphoton.2010.259}{\bibinfo {journal} {Nat. Photonics}}\ }%
  \textbf{\bibinfo {volume} {5}},\ \bibinfo {pages} {31} (\bibinfo {year}
  {2010})%
  \bibAnnoteFile{NoStop}{Kampfrath2010}%
\bibitem{Fleischer2011}%
  \BibitemOpen
  \bibfield{author}{%
  \bibinfo {author} {\bibfnamefont{S.}~\bibnamefont{Fleischer}}, \bibinfo
  {author} {\bibfnamefont{Y.}~\bibnamefont{Zhou}}, \bibinfo {author}
  {\bibfnamefont{R.~W.}\ \bibnamefont{Field}},\ and\ \bibinfo {author}
  {\bibfnamefont{K.~A.}\ \bibnamefont{Nelson}},\ }%
  \bibfield{journal}{%
  \Doi{10.1103/PhysRevLett.107.163603}{\bibinfo {journal} {Phys. Rev. Lett.}}\
  }%
  \textbf{\bibinfo {volume} {107}},\ \bibinfo {pages} {163603} (\bibinfo {year}
  {2011})%
  \bibAnnoteFile{NoStop}{Fleischer2011}%
\bibitem{Daigle2012}%
  \BibitemOpen
  \bibfield{author}{%
  \bibinfo {author} {\bibnamefont{{J. F. Daigle}}}, \bibinfo {author}
  {\bibfnamefont{F.}~\bibnamefont{Th\'{e}berge}}, \bibinfo {author}
  {\bibfnamefont{M.}~\bibnamefont{Henriksson}}, \bibinfo {author}
  {\bibnamefont{{T. J. Wang}}}, \bibinfo {author}
  {\bibfnamefont{S.}~\bibnamefont{Yuan}}, \bibinfo {author}
  {\bibfnamefont{M.}~\bibnamefont{Ch\^{a}teauneuf}}, \bibinfo {author}
  {\bibfnamefont{J.}~\bibnamefont{Dubois}}, \bibinfo {author}
  {\bibfnamefont{M.}~\bibnamefont{Pich\'{e}}},\ and\ \bibinfo {author}
  {\bibnamefont{{S. L. Chin}}},\ }%
  \bibfield{journal}{%
  \Doi{10.1364/OE.20.006825}{\bibinfo {journal} {Opt. Express}}\ }%
  \textbf{\bibinfo {volume} {20}},\ \bibinfo {pages} {6825} (\bibinfo {year}
  {2012})%
  \bibAnnoteFile{NoStop}{Daigle2012}%
\bibitem{Lu2006}%
  \BibitemOpen
  \bibfield{author}{%
  \bibinfo {author} {\bibfnamefont{M.}~\bibnamefont{Lu}}, \bibinfo {author}
  {\bibfnamefont{J.}~\bibnamefont{Shen}}, \bibinfo {author}
  {\bibfnamefont{N.}~\bibnamefont{Li}}, \bibinfo {author}
  {\bibfnamefont{Y.}~\bibnamefont{Zhang}}, \bibinfo {author}
  {\bibfnamefont{C.}~\bibnamefont{Zhang}}, \bibinfo {author}
  {\bibfnamefont{L.}~\bibnamefont{Liang}},\ and\ \bibinfo {author}
  {\bibfnamefont{X.}~\bibnamefont{Xu}},\ }%
  \bibfield{journal}{%
  \Doi{10.1063/1.2388041}{\bibinfo {journal} {J. Appl. Phys.}}\ }%
  \textbf{\bibinfo {volume} {100}},\ \bibinfo {pages} {103104} (\bibinfo {year}
  {2006})%
  \bibAnnoteFile{NoStop}{Lu2006}%
\bibitem{Cook2000}%
  \BibitemOpen
  \bibfield{author}{%
  \bibinfo {author} {\bibfnamefont{D.~J.}\ \bibnamefont{Cook}}\ and\ \bibinfo
  {author} {\bibfnamefont{R.~M.}\ \bibnamefont{Hochstrasser}},\ }%
  \bibfield{journal}{%
  \Doi{10.1364/OL.25.001210}{\bibinfo {journal} {Opt. Lett.}}\ }%
  \textbf{\bibinfo {volume} {25}},\ \bibinfo {pages} {1210} (\bibinfo {year}
  {2000})%
  \bibAnnoteFile{NoStop}{Cook2000}%
\bibitem{Oh2013}%
  \BibitemOpen
  \bibfield{author}{%
  \bibinfo {author} {\bibfnamefont{T.~I.}\ \bibnamefont{Oh}}, \bibinfo {author}
  {\bibfnamefont{Y.~S.}\ \bibnamefont{You}}, \bibinfo {author}
  {\bibfnamefont{N.}~\bibnamefont{Jhajj}}, \bibinfo {author}
  {\bibfnamefont{E.~W.}\ \bibnamefont{Rosenthal}}, \bibinfo {author}
  {\bibfnamefont{H.~M.}\ \bibnamefont{Milchberg}},\ and\ \bibinfo {author}
  {\bibfnamefont{K.~Y.}\ \bibnamefont{Kim}},\ }%
  \bibfield{journal}{%
  \Doi{10.1063/1.4807790}{\bibinfo {journal} {Appl. Phys. Lett.}}\ }%
  \textbf{\bibinfo {volume} {102}},\ \bibinfo {pages} {201113} (\bibinfo {year}
  {2013})%
  \bibAnnoteFile{NoStop}{Oh2013}%
\bibitem{Kim2007}%
  \BibitemOpen
  \bibfield{author}{%
  \bibinfo {author} {\bibfnamefont{K.~Y.}\ \bibnamefont{Kim}}, \bibinfo
  {author} {\bibfnamefont{J.~H.}\ \bibnamefont{Glownia}}, \bibinfo {author}
  {\bibfnamefont{A.~J.}\ \bibnamefont{Taylor}},\ and\ \bibinfo {author}
  {\bibfnamefont{G.}~\bibnamefont{Rodriguez}},\ }%
  \bibfield{journal}{%
  \Doi{10.1364/OE.15.004577}{\bibinfo {journal} {Opt. Express}}\ }%
  \textbf{\bibinfo {volume} {15}},\ \bibinfo {pages} {4577} (\bibinfo {year}
  {2007})%
  \bibAnnoteFile{NoStop}{Kim2007}%
\bibitem{Kim2009}%
  \BibitemOpen
  \bibfield{author}{%
  \bibinfo {author} {\bibfnamefont{K.~Y.}\ \bibnamefont{Kim}},\ }%
  \bibfield{journal}{%
  \Doi{10.1063/1.3134422}{\bibinfo {journal} {Phys. Plasmas}}\ }%
  \textbf{\bibinfo {volume} {16}},\ \bibinfo {pages} {056706} (\bibinfo {year}
  {2009})%
  \bibAnnoteFile{NoStop}{Kim2009}%
\bibitem{Berge2013}%
  \BibitemOpen
  \bibfield{author}{%
  \bibinfo {author} {\bibfnamefont{L.}~\bibnamefont{Berg\'{e}}}, \bibinfo
  {author} {\bibfnamefont{S.}~\bibnamefont{Skupin}}, \bibinfo {author}
  {\bibfnamefont{C.}~\bibnamefont{K\"{o}hler}}, \bibinfo {author}
  {\bibfnamefont{I.}~\bibnamefont{Babushkin}},\ and\ \bibinfo {author}
  {\bibfnamefont{J.}~\bibnamefont{Herrmann}},\ }%
  \bibfield{journal}{%
  \Doi{10.1103/PhysRevLett.110.073901}{\bibinfo {journal} {Phys. Rev. Lett.}}\
  }%
  \textbf{\bibinfo {volume} {110}},\ \bibinfo {pages} {073901} (\bibinfo {year}
  {2013})%
  \bibAnnoteFile{NoStop}{Berge2013}%
\bibitem{You2012}%
  \BibitemOpen
  \bibfield{author}{%
  \bibinfo {author} {\bibfnamefont{Y.~S.}\ \bibnamefont{You}}, \bibinfo
  {author} {\bibfnamefont{T.~I.}\ \bibnamefont{Oh}},\ and\ \bibinfo {author}
  {\bibfnamefont{K.~Y.}\ \bibnamefont{Kim}},\ }%
  \bibfield{journal}{%
  \Doi{10.1103/PhysRevLett.109.183902}{\bibinfo {journal} {Phys. Rev. Lett.}}\
  }%
  \textbf{\bibinfo {volume} {109}},\ \bibinfo {pages} {183902} (\bibinfo {year}
  {2012})%
  \bibAnnoteFile{NoStop}{You2012}%
\bibitem{Xu2013}%
  \BibitemOpen
  \bibfield{author}{%
  \bibinfo {author} {\bibfnamefont{R.}~\bibnamefont{Xu}}, \bibinfo {author}
  {\bibfnamefont{Y.}~\bibnamefont{Bai}}, \bibinfo {author}
  {\bibfnamefont{L.}~\bibnamefont{Song}}, \bibinfo {author}
  {\bibfnamefont{P.}~\bibnamefont{Liu}}, \bibinfo {author}
  {\bibfnamefont{R.}~\bibnamefont{Li}},\ and\ \bibinfo {author}
  {\bibfnamefont{Z.}~\bibnamefont{Xu}},\ }%
  \bibfield{journal}{%
  \Doi{10.1063/1.4817975}{\bibinfo {journal} {Appl. Phys. Lett.}}\ }%
  \textbf{\bibinfo {volume} {103}},\ \bibinfo {pages} {061111} (\bibinfo {year}
  {2013})%
  \bibAnnoteFile{NoStop}{Xu2013}%
\bibitem{Dai2011}%
  \BibitemOpen
  \bibfield{author}{%
  \bibinfo {author} {\bibfnamefont{H.}~\bibnamefont{Dai}}\ and\ \bibinfo
  {author} {\bibfnamefont{J.}~\bibnamefont{Liu}},\ }%
  \bibfield{journal}{%
  \Doi{10.1088/2040-8978/13/5/055201}{\bibinfo {journal} {J. Opt.}}\ }%
  \textbf{\bibinfo {volume} {13}},\ \bibinfo {pages} {055201} (\bibinfo {year}
  {2011})%
  \bibAnnoteFile{NoStop}{Dai2011}%
\bibitem{Auston1984}%
  \BibitemOpen
  \bibfield{author}{%
  \bibinfo {author} {\bibfnamefont{D.~H.}\ \bibnamefont{Auston}}, \bibinfo
  {author} {\bibfnamefont{K.~P.}\ \bibnamefont{Cheung}}, \bibinfo {author}
  {\bibfnamefont{J.~A.}\ \bibnamefont{Valdmanis}},\ and\ \bibinfo {author}
  {\bibfnamefont{D.~A.}\ \bibnamefont{Kleinman}},\ }%
  \bibfield{journal}{%
  \bibinfo {journal} {Phys. Rev. Lett.}\ }%
  \textbf{\bibinfo {volume} {53}},\ \bibinfo {pages} {1555} (\bibinfo {year}
  {1984})%
  \bibAnnoteFile{NoStop}{Auston1984}%
\bibitem{Amico2008}%
  \BibitemOpen
  \bibfield{author}{%
  \bibinfo {author} {\bibfnamefont{C.}~\bibnamefont{{D'Amico}}}, \bibinfo
  {author} {\bibfnamefont{A.}~\bibnamefont{Houard}}, \bibinfo {author}
  {\bibfnamefont{S.}~\bibnamefont{Akturk}}, \bibinfo {author}
  {\bibfnamefont{Y.}~\bibnamefont{Liu}}, \bibinfo {author}
  {\bibfnamefont{J.}~\bibnamefont{{Le Bloas}}}, \bibinfo {author}
  {\bibfnamefont{M.}~\bibnamefont{Franco}}, \bibinfo {author}
  {\bibfnamefont{B.}~\bibnamefont{Prade}}, \bibinfo {author}
  {\bibfnamefont{A.}~\bibnamefont{Couairon}}, \bibinfo {author}
  {\bibfnamefont{V.~T.}\ \bibnamefont{Tikhonchuk}},\ and\ \bibinfo {author}
  {\bibfnamefont{A.}~\bibnamefont{Mysyrowicz}},\ }%
  \bibfield{journal}{%
  \Doi{10.1088/1367-2630/10/1/013015}{\bibinfo {journal} {New J. Phys.}}\ }%
  \textbf{\bibinfo {volume} {10}},\ \bibinfo {pages} {013015} (\bibinfo {year}
  {2008})%
  \bibAnnoteFile{NoStop}{Amico2008}%
\bibitem{Kohler2011}%
  \BibitemOpen
  \bibfield{author}{%
  \bibinfo {author} {\bibfnamefont{C.}~\bibnamefont{K\"{o}hler}}, \bibinfo
  {author} {\bibfnamefont{E.}~\bibnamefont{Cabrera-Granado}}, \bibinfo {author}
  {\bibfnamefont{I.}~\bibnamefont{Babushkin}}, \bibinfo {author}
  {\bibfnamefont{L.}~\bibnamefont{Berg\'{e}}}, \bibinfo {author}
  {\bibfnamefont{J.}~\bibnamefont{Herrmann}},\ and\ \bibinfo {author}
  {\bibfnamefont{S.}~\bibnamefont{Skupin}},\ }%
  \bibfield{journal}{%
  \Doi{10.1364/OL.36.003166}{\bibinfo {journal} {Opt. Lett.}}\ }%
  \textbf{\bibinfo {volume} {36}},\ \bibinfo {pages} {3166} (\bibinfo {year}
  {2011})%
  \bibAnnoteFile{NoStop}{Kohler2011}%
\bibitem{Kolesik2004}%
  \BibitemOpen
  \bibfield{author}{%
  \bibinfo {author} {\bibfnamefont{M.}~\bibnamefont{Kolesik}}\ and\ \bibinfo
  {author} {\bibfnamefont{J.~V.}\ \bibnamefont{Moloney}},\ }%
  \bibfield{journal}{%
  \Doi{10.1103/PhysRevE.70.036604}{\bibinfo {journal} {Phys. Rev. E}}\ }%
  \textbf{\bibinfo {volume} {70}},\ \bibinfo {pages} {036604} (\bibinfo {year}
  {2004})%
  \bibAnnoteFile{NoStop}{Kolesik2004}%
\bibitem{Peck1966}%
  \BibitemOpen
  \bibfield{author}{%
  \bibinfo {author} {\bibfnamefont{E.~R.}\ \bibnamefont{Peck}}\ and\ \bibinfo
  {author} {\bibfnamefont{B.~N.}\ \bibnamefont{Khanna}},\ }%
  \bibfield{journal}{%
  \bibinfo {journal} {J. Opt. Soc. Am.}\ }%
  \textbf{\bibinfo {volume} {56}},\ \bibinfo {pages} {1059} (\bibinfo {year}
  {1966})%
  \bibAnnoteFile{NoStop}{Peck1966}%
\bibitem{Lu2012}%
  \BibitemOpen
  \bibfield{author}{%
  \bibinfo {author} {\bibfnamefont{X.}~\bibnamefont{Lu}}\ and\ \bibinfo
  {author} {\bibfnamefont{X.-C.}\ \bibnamefont{Zhang}},\ }%
  \bibfield{journal}{%
  \Doi{10.1103/PhysRevLett.108.123903}{\bibinfo {journal} {Phys. Rev. Lett.}}\
  }%
  \textbf{\bibinfo {volume} {108}},\ \bibinfo {pages} {123903} (\bibinfo {year}
  {2012})%
  \bibAnnoteFile{NoStop}{Lu2012}%
\bibitem{Wahlstrand2012}%
  \BibitemOpen
  \bibfield{author}{%
  \bibinfo {author} {\bibfnamefont{J.}~\bibnamefont{Wahlstrand}}, \bibinfo
  {author} {\bibfnamefont{Y.-H.}\ \bibnamefont{Cheng}},\ and\ \bibinfo {author}
  {\bibfnamefont{H.~M.}\ \bibnamefont{Milchberg}},\ }%
  \bibfield{journal}{%
  \Doi{10.1103/PhysRevA.85.043820}{\bibinfo {journal} {Phys. Rev. A}}\ }%
  \textbf{\bibinfo {volume} {85}},\ \bibinfo {pages} {043820} (\bibinfo {year}
  {2012})%
  \bibAnnoteFile{NoStop}{Wahlstrand2012}%
\bibitem{Palastro2012}%
  \BibitemOpen
  \bibfield{author}{%
  \bibinfo {author} {\bibfnamefont{J.~P.}\ \bibnamefont{Palastro}}, \bibinfo
  {author} {\bibfnamefont{T.~M.}\ \bibnamefont{Antonsen}},\ and\ \bibinfo
  {author} {\bibfnamefont{H.~M.}\ \bibnamefont{Milchberg}},\ }%
  \bibfield{journal}{%
  \Doi{10.1103/PhysRevA.86.033834}{\bibinfo {journal} {Phys. Rev. A}}\ }%
  \textbf{\bibinfo {volume} {86}},\ \bibinfo {pages} {033834} (\bibinfo {year}
  {2012})%
  \bibAnnoteFile{NoStop}{Palastro2012}%
\bibitem{Sprangle2004}%
  \BibitemOpen
  \bibfield{author}{%
  \bibinfo {author} {\bibfnamefont{P.}~\bibnamefont{Sprangle}}, \bibinfo
  {author} {\bibfnamefont{J.}~\bibnamefont{Pe\~{n}ano}}, \bibinfo {author}
  {\bibfnamefont{B.}~\bibnamefont{Hafizi}},\ and\ \bibinfo {author}
  {\bibfnamefont{C.}~\bibnamefont{Kapetanakos}},\ }%
  \bibfield{journal}{%
  \bibinfo {journal} {Phys. Rev. E}\ }%
  \textbf{\bibinfo {volume} {69}},\ \bibinfo {pages} {066415} (\bibinfo {year}
  {2004})%
  \bibAnnoteFile{NoStop}{Sprangle2004}%
\bibitem{Popruzhenko2008}%
  \BibitemOpen
  \bibfield{author}{%
  \bibinfo {author} {\bibfnamefont{S.}~\bibnamefont{Popruzhenko}}, \bibinfo
  {author} {\bibfnamefont{V.}~\bibnamefont{Mur}}, \bibinfo {author}
  {\bibfnamefont{V.}~\bibnamefont{Popov}},\ and\ \bibinfo {author}
  {\bibfnamefont{D.}~\bibnamefont{Bauer}},\ }%
  \bibfield{journal}{%
  \Doi{10.1103/PhysRevLett.101.193003}{\bibinfo {journal} {Phys. Rev. Lett.}}\
  }%
  \textbf{\bibinfo {volume} {101}},\ \bibinfo {pages} {93003} (\bibinfo {year}
  {2008})%
  \bibAnnoteFile{NoStop}{Popruzhenko2008}%
\bibitem{Babushkin2010}%
  \BibitemOpen
  \bibfield{author}{%
  \bibinfo {author} {\bibfnamefont{I.}~\bibnamefont{Babushkin}}, \bibinfo
  {author} {\bibfnamefont{W.}~\bibnamefont{Kuehn}}, \bibinfo {author}
  {\bibfnamefont{C.}~\bibnamefont{K\"{o}hler}}, \bibinfo {author}
  {\bibfnamefont{S.}~\bibnamefont{Skupin}}, \bibinfo {author}
  {\bibfnamefont{L.}~\bibnamefont{Berg\'{e}}}, \bibinfo {author}
  {\bibfnamefont{K.}~\bibnamefont{Reimann}}, \bibinfo {author}
  {\bibfnamefont{M.}~\bibnamefont{Woerner}}, \bibinfo {author}
  {\bibfnamefont{J.}~\bibnamefont{Herrmann}},\ and\ \bibinfo {author}
  {\bibfnamefont{T.}~\bibnamefont{Elsaesser}},\ }%
  \bibfield{journal}{%
  \Doi{10.1103/PhysRevLett.105.053903}{\bibinfo {journal} {Phys. Rev. Lett.}}\
  }%
  \textbf{\bibinfo {volume} {105}},\ \bibinfo {pages} {053903} (\bibinfo {year}
  {2010})%
  \bibAnnoteFile{NoStop}{Babushkin2010}%
\bibitem{Sprangle2002}%
  \BibitemOpen
  \bibfield{author}{%
  \bibinfo {author} {\bibfnamefont{P.}~\bibnamefont{Sprangle}}, \bibinfo
  {author} {\bibfnamefont{J.}~\bibnamefont{Pe\~{n}ano}},\ and\ \bibinfo
  {author} {\bibfnamefont{B.}~\bibnamefont{Hafizi}},\ }%
  \bibfield{journal}{%
  \Doi{10.1103/PhysRevE.66.046418}{\bibinfo {journal} {Phys. Rev. E}}\ }%
  \textbf{\bibinfo {volume} {66}},\ \bibinfo {pages} {046418} (\bibinfo {year}
  {2002})%
  \bibAnnoteFile{NoStop}{Sprangle2002}%
\bibitem{Rodriguez2010}%
  \BibitemOpen
  \bibfield{author}{%
  \bibinfo {author} {\bibfnamefont{G.}~\bibnamefont{Rodriguez}}\ and\ \bibinfo
  {author} {\bibfnamefont{G.~L.}\ \bibnamefont{Dakovski}},\ }%
  \bibfield{journal}{%
  \bibinfo {journal} {Opt. Express}\ }%
  \textbf{\bibinfo {volume} {18}},\ \bibinfo {pages} {15130} (\bibinfo {year}
  {2010})%
  \bibAnnoteFile{NoStop}{Rodriguez2010}%
\bibitem{Blank2013}%
  \BibitemOpen
  \bibfield{author}{%
  \bibinfo {author} {\bibfnamefont{V.}~\bibnamefont{Blank}}, \bibinfo {author}
  {\bibfnamefont{M.~D.}\ \bibnamefont{Thomson}},\ and\ \bibinfo {author}
  {\bibfnamefont{H.~G.}\ \bibnamefont{Roskos}},\ }%
  \bibfield{journal}{%
  \Doi{10.1088/1367-2630/15/7/075023}{\bibinfo {journal} {New J. Phys.}}\ }%
  \textbf{\bibinfo {volume} {15}},\ \bibinfo {pages} {075023} (\bibinfo {year}
  {2013})%
  \bibAnnoteFile{NoStop}{Blank2013}%
\bibitem{Keldysh1965}%
  \BibitemOpen
  \bibfield{author}{%
  \bibinfo {author} {\bibfnamefont{L.~V.}\ \bibnamefont{Keldysh}},\ }%
  \bibfield{journal}{%
  \bibinfo {journal} {Sov. Phys. JETP}\ }%
  \textbf{\bibinfo {volume} {20}},\ \bibinfo {pages} {1307} (\bibinfo {year}
  {1965})%
  \bibAnnoteFile{NoStop}{Keldysh1965}%
\bibitem{Popov2004}%
  \BibitemOpen
  \bibfield{author}{%
  \bibinfo {author} {\bibfnamefont{V.~S.}\ \bibnamefont{Popov}},\ }%
  \bibfield{journal}{%
  \Doi{10.1070/PU2004v047n09ABEH001812}{\bibinfo {journal} {Physics-Uspekhi}}\
  }%
  \textbf{\bibinfo {volume} {47}},\ \bibinfo {pages} {855} (\bibinfo {year}
  {2004})%
  \bibAnnoteFile{NoStop}{Popov2004}%
\bibitem{Perelomov1966}%
  \BibitemOpen
  \bibfield{author}{%
  \bibinfo {author} {\bibfnamefont{A.~M.}\ \bibnamefont{Perelomov}}, \bibinfo
  {author} {\bibfnamefont{V.~S.}\ \bibnamefont{Popov}},\ and\ \bibinfo {author}
  {\bibfnamefont{M.~V.}\ \bibnamefont{Terentev}},\ }%
  \bibfield{journal}{%
  \bibinfo {journal} {Sov. Phys. JETP}\ }%
  \textbf{\bibinfo {volume} {23}},\ \bibinfo {pages} {924} (\bibinfo {year}
  {1966})%
  \bibAnnoteFile{NoStop}{Perelomov1966}%
\bibitem{COUAIRON2007}%
  \BibitemOpen
  \bibfield{author}{%
  \bibinfo {author} {\bibfnamefont{A.}~\bibnamefont{Couairon}}\ and\ \bibinfo
  {author} {\bibfnamefont{A.}~\bibnamefont{Mysyrowicz}},\ }%
  \bibfield{journal}{%
  \Doi{10.1016/j.physrep.2006.12.005}{\bibinfo {journal} {Physics Reports}}\ }%
  \textbf{\bibinfo {volume} {441}},\ \bibinfo {pages} {47} (\bibinfo {year}
  {2007})%
  \bibAnnoteFile{NoStop}{COUAIRON2007}%
\bibitem{Larochelle1998}%
  \BibitemOpen
  \bibfield{author}{%
  \bibinfo {author} {\bibfnamefont{S.~F.~J.}\ \bibnamefont{Larochelle}},
  \bibinfo {author} {\bibfnamefont{A.}~\bibnamefont{Talebpour}},\ and\ \bibinfo
  {author} {\bibfnamefont{S.~L.}\ \bibnamefont{Chin}},\ }%
  \bibfield{journal}{%
  \Doi{10.1088/0953-4075/31/6/009}{\bibinfo {journal} {J. Phys. B}}\ }%
  \textbf{\bibinfo {volume} {31}},\ \bibinfo {pages} {1215} (\bibinfo {year}
  {1998})%
  \bibAnnoteFile{NoStop}{Larochelle1998}%
\bibitem{Talebpour1999}%
  \BibitemOpen
  \bibfield{author}{%
  \bibinfo {author} {\bibfnamefont{A.}~\bibnamefont{Talebpour}},\ }%
  \bibfield{journal}{%
  \Doi{10.1016/S0030-4018(99)00113-3}{\bibinfo {journal} {Opt. Commun.}}\ }%
  \textbf{\bibinfo {volume} {163}},\ \bibinfo {pages} {29} (\bibinfo {year}
  {1999})%
  \bibAnnoteFile{NoStop}{Talebpour1999}%
\bibitem{Kotelnikov2011}%
  \BibitemOpen
  \bibfield{author}{%
  \bibinfo {author} {\bibfnamefont{I.~A.}\ \bibnamefont{Kotelnikov}}, \bibinfo
  {author} {\bibfnamefont{A.~V.}\ \bibnamefont{Borodin}},\ and\ \bibinfo
  {author} {\bibfnamefont{A.~P.}\ \bibnamefont{Shkurinov}},\ }%
  \bibfield{journal}{%
  \Doi{10.1134/S1063776111050049}{\bibinfo {journal} {Sov. Phys. JEPT}}\ }%
  \textbf{\bibinfo {volume} {112}},\ \bibinfo {pages} {946} (\bibinfo {year}
  {2011})%
  \bibAnnoteFile{NoStop}{Kotelnikov2011}%
\end{thebibliography}%
\end{document}